# Base Station Selection And Task Offloading Of The Mobile Edge Computing System

Ruan Yanjiao

College of Information Science and Engineering, Hunan University, 410082,ChangSha,China

ryj@hnu.edu.cn

**Abstract**.

Based on the two decision variables, service location and base station selection, construct a computational model of the switching delay, communication delay, and queuing delay patterns of a mobile edge computing system in each time horizon; minimize the non-switching latency to obtain the service deployment and base station selection decision at the initial time horizon; compute the switching latency and non-switching latency of the current time horizon based on the decision of the previous time horizon, and determine the current time horizon based on the principle of larger non-switching latency tolerance. If the service is not reallocated, the decision for the current time slot remains the same as the decision for the previous time slot; if the service is reallocated, the non-reach time of the current time slot is minimized and the service allocation and base station selection decision for the current time slot is obtained; the service allocation and base station selection decisions for all time slots are iteratively computed. This paper provides low-latency processing and quality of service for users to satisfy the randomness of user movement.

## 1 Introduction

In recent years, the rapid development of mobile devices, represented by new services such as virtual/augmented reality, autonomous driving and telematics, has increased the demands on mobile communications. Mobile edge computing has emerged to support low latency and high reliability applications [1].

The foundation of mobile edge computing is the placement of storage and compute resources near the edge of the network at the user, thereby reducing latency and providing quality of service to the user. In the case of Extreme Smart Grids, the edge cloud of the network consists mainly of micro data centers, with each end of the network configured with an access point, also called a base station [2]. Users reduce latency by choosing to access services from a base station located close to the edge network rather than a remote cloud center.

User mobility is often an issue in extreme smart grids. A common solution is service relocation. Typically, services are moved closer to the user to reduce communication delay. In other words, after the subscriber moves, the MEC system makes the following decisions to achieve the goal of optimizing the subscriber's service quality: first, whether to change the location of the service requested by the mobile subscriber; then, if and when the operating unit decides to change, and to which edge cloud in the network the service is transferred [3].

If a user chooses to access an edge cloud that has a long queue of tasks to process at the base station, this will cause a large queuing delay for the current user [4]. Obviously, if the edge cloud has services for multiple users at the same time, the base station will be heavily loaded, which will also affect the task execution time. Therefore, optimizing the selection of appropriate base station and service placement for the subscriber can improve the quality of service for the subscriber.

## 2 System Model

The Edge Intelligence Network has a total of three clouds and one user. The server cloud consists of base stations and MEC servers. For example, taking user 1 as an example, there are base stations in edge clouds 1, 2 that can be selected to access the service range and communicate with each other over a wireless channel. The service requested by user 1 can be deployed in any edge cloud, typically selecting an edge cloud with lower load, for example edge cloud 3. The edge clouds communicate with each other via a communication link, for example between base station 1 and base station 2 via communication link 1. At this point, the queue of Edge 1 and Edge 2 base stations is long, which affects the delay of the queued tasks; at the same time, the load of the base stations is too high; Edge 3 base stations have no queue and are lightly loaded, but the users are not in the range of Edge 3. The optimization problem is performed by the following steps [5].

1) User 1 calculates the delay based on the current switching delay and non-switching delay information, and decides whether the service should be moved based on the principle of tolerating as much non-switching delay as possible.

2) Deploy the service in the edge cloud 3 based on the outcome of whether the service should be relocated and the principle of minimizing the non-localization delay.

Case 1: The service is not migrated, which is consistent with the decision made in the previous period.

Case 2: The service is migrated, the constraints are relaxed separately, and the fractional results of the service deployment decision and the base station selection decision are obtained using a linear programming solver, and the decision is set to 1 according to the fractional result of the deployment decision with probability 0 otherwise.

(3) Based on the multicast delay minimization principle and the service deployment result, a network access point of user 1, i.e., base station 1, is selected.

(4) The user uses the communication link between base station 1 and base station 3 to obtain the required services deployed in the edge cloud 3.

Said mobile edge computing system comprises a cloud of edge computing devices, each cloud of edge computing devices comprising a base station and a mobile edge server, said base station being used to access and download services from mobile user devices to the local mobile edge server or to other clouds of edge computing devices via a communication link [6]; more particularly, this may be expressed as follows.

$\mathcal{M} = \{0,1,\ldots M\}$ represents the collection of $\mathcal{M}$ edge computing clouds; $\mathcal{N} = \{0,1,\ldots N\}$ represents a collection of N mobile users; The time-slot division of the system is expressed as

$\mathcal{T} = \{0,1,\ldots \tau\};$ In the time slot $t \in \mathcal{T}$, The set of base stations to which the user $k \in \mathcal{N}$ can access is expressed as $\phi_k(t)$; Service placement decision variable $x_{ik}(t) \in \{0,1\}$ Represents the time slot t, Whether the service required for the user $k \in \mathcal{N}$ are placed in the edge cloud $i \in \mathcal{M}$, among, The 0 is no, 1 is for the yes, $x(t)$ A vector representation of the decision variable $x_{ik}(t)$ for the service; Base station selection decision variable $y_{jk}(t) \in \{0,1\}$ represents the time slot t, Whether the user $k \in \mathcal{N}$ selects the base station $j \in \phi_k(t)$ within the access service range, among, The 0 is no, 1 is for the yes, $y(t)$ is the vector representation of the decision variable $y_{jk}(t)$.

Delay Computation Model Construction: construct a model to compute the switching delay, communication delay, and queuing delay of a mobile edge computing system for each time interval based on two decision variables: Service Deployment Decision and Base Station Selection Decision [7]; obtaining a model to compute the total delay from the sum of the switching delay, communication delay and queuing delay and obtaining a model to compute the non-switching delay from the sum of the communication delay and queuing delay.

Task processing delay in mobile edge computing systems includes queuing, transition delay, and communication delay, each of which is explained below.

**Transition delay**: This occurs when users move across the coverage area of multiple base stations. Dynamic deployment or migration of services is unavoidable to ensure quality of service for users. From time to time, the total switching delay for all users is expressed as follows:

$$T^s(x(t), x(t-1)) = \sum_{k \in \mathcal{N}} \sum_{i \in \mathcal{M}} s_i [x_{ik}(t) - x_{ik}(t-1)]^+$$

Where $s_i(t)$ represents the user service resource requirements on the edge cloud $i \in \mathcal{M}$, $[x_{ik}(t) - x_{ik}(t-1)]^+ = \max\{x_{ik}(t) - x_{ik}(t-1), 0\}$.

**Queuing delay**: occurs when connecting user equipment to the base station. In the case of a base station, the number of subscribers connected to it changes over time. If too many users are connected to the base station, or even if they exceed the load of the base station, queuing delay will affect the quality of service for users. The model for calculating queuing delay can be expressed as follows [7].

$$T^q(y(t)) = \sum_{k \in \mathcal{N}} \sum_{i \in \mathcal{M}} y_{jk}(t) \frac{1}{C_j - \sum_{k \in \mathcal{N}} c_k(t) y_{jk}(t)}$$

Where $C_j$ is the capacity of the base station in the edge intelligent network, and $c_k(t)$ is the service demand of the time-slot users.

**Communication delay**: occurs in the process of user devices accessing services through the edge cloud. In the mobile edge computing model of the present invention, the location of the services is independent and need not necessarily be located in the cloud associated with the base station accessed by the user. The communication time delay corresponding to the base station selection decision and the service deployment decision is expressed as:

$$T^c(x(t), y(t)) = \sum_{k \in \mathcal{N}} \sum_{i \in \mathcal{M}} \sum_{j \in \phi_k(t)} y_{jk}(t) x_{ik}(t) l_{ij}(t)$$

Where $l_{ij}(t)$ is the transmission delay of base station $j \in \phi_k(t)$ to edge cloud $i \in \mathcal{M}$.

Therefore, the overall completion time of the task can be expressed as:

$$\sum_{t=1}^{\tau} T(x(t), y(t)) = \sum_{t=1}^{\tau} (T^q(y(t)) + T^s(x(t), x(t-1)) + T^c(x(t), y(t)))$$

Where $T(x(t), y(t))$ is the total system delay of the time slot.

The above queuing delay and communication delay are non-switching delay, and they only depend on the system information of the current time slot, so the non-switching delay can be expressed as:

$$T^{ns}(x(t), y(t)) = T^q(y(t)) + T^c(x(t), y(t))$$

Where $T^{ns}(x(t), y(t))$ is the non-switching delay of the time slot.

### 3 Problem Description

The start time is fixed, and the non-transient delay is used as an optimization problem to obtain a service deployment decision and a base station selection decision at the start time by minimizing the non-transient delay at the start time.

According to the non-switching delay $T^{ns}(x(t), y(t)) = T^q(y(t)) + T^c(x(t), y(t))$, it is used as an optimization problem to obtain the service placement decision and base station selection decision at the initial moment $t_1$.

Among them, to achieve the above optimization objectives, the following constraints exist:

The first constraint indicates that each service can only be deployed on one cloud edge in the smart grid at any given time t: $\sum_{i \in \mathcal{M}} x_{ik}(t) = 1, \forall k \in \mathcal{N}$;

The second constraint indicates that the total number of services in each cloud must not exceed the capacity constraint, where $S_i$ the total number of services in the cloud $i \in$ is indicated: $\sum_{k \in \mathcal{N}} s_k(t) x_{ik}(t) \leq S_i$, $\forall i \in \mathcal{M}$ ;

The third constraint indicates whether the service required for the user's task $k \in \mathcal{N}$ is deployed on a cloud $i \in \mathcal{M}$, where 0 means no and 1 means yes: $x_{ik}(t) \in \{0,1\}, \forall i \in \mathcal{M}, \forall k \in \mathcal{N}$;

The fourth constraint specifies that each user may use only one base station in the service area in each time slot: $\sum_{j \in \phi_k(t)} y_{jk}(t) = 1$, $\forall k \in \mathcal{N}$ [8];

The fifth constraint specifies whether the user $k \in \mathcal{N}$ chooses to use a base station $j \in \phi_k(t)$ in the service area, where 0 means "no" and 1 means "yes": $y_{jk}(t) \in \{0,1\}, \forall j \in \phi_k(t), \forall k \in \mathcal{N}$;

The sixth constraint specifies that the user's demand for service must not exceed the resource limit of the base station: $\sum_{k \in \mathcal{N}} c_k(t) y_{jk}(t) \leq C_j$, $\forall j \in \phi_k(t)$.

### 4 Method Introduction

Based on the service deployment decision and the base station selection decision for the previous time slot, the time slot switching delay and the non-switching delay are calculated, and the amount of non-switching delay from the last switching service to the current time slot is considered; the tolerance principle of the larger non-switching delay is used to determine whether the time slot service is transferred [8].

The threshold of the tolerance factor is predetermined depending on the type of offloading task as β, and the time slot switchover delay is recorded as T1 and the amount of non-switchover delay from the last service switchover to the current time slot switchover is recorded as T2. Based on the principle of tolerating a larger non-switchover delay, the method for determining whether the service should be migrated for each time slot is: compare T2 and $\beta * T1$, if $T2 < \beta * T1$, decide not to migrate, otherwise decide to migrate the service.

If the decision is made not to migrate the service, the decision to place the service and the decision to select a base station remain the same for that timeslot as for the previous timeslot $t - 1$.

If a decision is made to transfer the service, the non-transfer delay is used as an optimization problem to obtain the service transfer decision and the time recording base station selection decision, minimizing the non-transfer delay [9].

In practice, the total handover delay should be used as the optimization problem, but since the migration service is currently defined, the handover delay is defined and the optimization problem based on the total handover delay can be reduced to an optimization problem based on the nonhandover delay [9].

When solving the optimization problem based on the constraints, the third and fifth constraints are respectively relaxed to obtain:

$$x_{ik}(t) \in [0,1], \forall i \in \mathcal{M}, \forall k \in \mathcal{N};$$

$$y_{jk}(t) \in [0,1], \forall j \in \varphi_k(t), \forall k \in \mathcal{N};$$

Then the score results of service placement decision and base station selection decision are obtained through linear planning solution;

Finally, the score result of the service placement decision is taken as the probability of the service placement decision is 1 [10]; and the ratio of the score result of the service placement decision and the score result of the base station selection decision is taken as the probability of the base station selection decision is 1, namely:

$$P\{\bar{x}_{ik}(t) = 1\} = x_{ik}(t); P\{\bar{x}_{ik}(t) = 0\} = 1 - x_{ik}(t);$$

$$P\{\bar{y}_{ik}(t) = 1\} = \frac{y_{jk}(t)}{x_{ik}(t)}; P\{\bar{x}_{ik}(t) = 0\} = 1 - \frac{y_{jk}(t)}{x_{ik}(t)};$$

Where $\bar{x}_{ik}(t)$ is the integer solution of the service placement decision, and $\bar{y}_{ik}(t)$ is the integer solution of the base station placement decision [11].

Iteratively calculates all time slots for station placement decisions and base station selection decisions [12].

## 5 Conclusion

Satisfying quality of service in mobile peripheral computing systems with limited or uncertain future information is a long-term optimization challenge. The present invention provides a method for base station selection and task offloading for mobile edge computing systems that decomposes the long-term problem into multiple individual release-time problems and minimizes the total time cost of user tasks in the system, achieving the optimization objective for each release-time problem. In each time slot, the invention first chooses to tolerate as much continuous delay as possible and decides whether the services should be switched; then, based on the result of the decision, decides to select a base station differently and then computes the feasible probability of determining the service deployment decision based on the result of the base station selection decision; the system decides base station selection and service deployment based on these two decisions, n The system then makes base station selection and service deployment decisions based on these two decisions, thereby minimizing task delay and significantly improving the quality of service for users.